\documentclass{JFM-FLM_Au}

\lefttitle{T.C.~Sykes and others}
\righttitle{Journal of Fluid Mechanics}

\usepackage{dcolumn}				
\usepackage{nicefrac}				
\usepackage{siunitx}				
\usepackage{xfrac}                  

\hypersetup{
    colorlinks,
    linkcolor=blue,
    citecolor=blue,
    urlcolor=blue
}

\DeclareSIUnit\px{pixels}
\DeclareSIUnit\stokes{St}

\newcommand{\Reyt}{\textit{Re}_t}
\newcommand{\We}{\mbox{\textit{We}}}
\newcommand{\Ca}{\mbox{\textit{Ca}}}
\newcommand{\tmustar}{t_\mu^\ast}
\newcommand{\ut}{u_t}
\newcommand{\un}{u_n}
\newcommand{\uo}{u_o}
\newcommand{\rn}{r_n}
\newcommand{\quantity}{\sqrt{\ut/\un}}

\title{Fast droplet impact onto slowly moving deep pools}

\author{Thomas C. Sykes\aff{1,}\aff{2}, Luke F.L. Alventosa\aff{3}, J. Rafael Castrej\'on-Pita\aff{4},\\ Radu Cimpeanu\aff{5}, Daniel M. Harris\aff{3}, \and Alfonso A. Castrej\'on-Pita\aff{2}}

\affiliation{\aff{1} School of Engineering, University of Warwick, Coventry CV4 7AL, UK
\aff{2} Department of Engineering Science, University of Oxford, Oxford OX1 3PJ, UK
\aff{3} School of Engineering, Brown University, Providence, RI 02912, USA
\aff{4} Department of Mechanical Engineering, University College London, London WC1E 7JE, UK
\aff{5} Mathematics Institute, University of Warwick, Coventry CV4 7AL, UK}

\corresau{Thomas C. Sykes, \email{thomas.sykes@warwick.ac.uk} and \email{t.c.sykes@outlook.com}}

\begin{document}
\maketitle

\begin{abstract}
    When a fast droplet impacts a pool, the resulting ejecta sheet dynamics determine the final impact outcome. At low Capillary numbers, the ejecta sheet remains separate from a deep static pool, whilst at higher viscosities it develops into a lamella. Here, we show that the common natural scenario of a slowly moving deep pool can change the upstream impact outcome, creating highly three-dimensional dynamics no longer characterised by a single descriptor. By considering how pool movement constrains the evolution of the ejecta sheet angle, we reach a length-scale invariant parameterisation for the upstream transition that holds for a wide range of fluids and impact conditions. Direct numerical simulations show similar dynamics for an equivalent oblique impact, indicating that the pool boundary layer does not play a decisive role for low pool-droplet speed ratios. Our results also provide insight into the physical mechanism that underpins pool impact outcomes more generally.
\end{abstract}

\section{Introduction}
    \label{sec:intro}

Most droplet impact research concerns normal collisions with a static substrate. However, in many natural and industrial processes, droplets impact obliquely or onto a moving substrate. This seemingly subtle distinction transforms droplet impact from the typical axisymmetric configuration to be inherently three-dimensional (3D). Example processes include raindrop splashing on the moving ocean that contributes to air–sea exchange \citep{Anguelova2021}, inkjet printing onto moving paper \citep{Detlef2022}, and crop spraying with numerous droplet impacts at arbitrary angles \citep{GielenPRF2017}. 

Most of the existing literature on moving substrates has considered dry surfaces, where horizontal surface motion can modify splashing thresholds \citep{Bird2009,HaoJiguang2017}, alter spreading factors \citep{Li2024moving}, and even generate new impact outcomes like boundary layer-driven ``aerodynamic rebound'' \citep{Stumpf2025}. Likewise, inclined static surfaces yield asymmetric crowns and reduce splashing propensity \citep{HaoPRL2019}.

When the substrate is a pool, either moving or with a droplet impacting obliquely, the literature is more limited and exploratory. \citet{Castrejon2016} identified distinct regimes (including a new `surfing' outcome) that depend on the droplet-pool speed ratio, but could not explore low pool speeds. Several studies have considered thin films that are either flowing \citep{Guo2025} or obliquely impacted \citep{Bao2025}. A few authors have considered oblique impact \citep{GielenPRF2017,Reijers2019}, and moving \citep{Gupta2020}, deep pools. However, all of these works focused on long time-scale dynamics such as crater evolution, crown structure, and the Worthington jet.

For static pools, following the observation that high-speed droplet impact leads to the formation of an ejecta sheet \citep{WeissEjecta,SiggiEjecta}, it is now well understood that the ejecta sheet dynamics on very short time scales determine impact outcomes \citep{Wang2023jfm}. On a deep pool, the ejecta sheet may collect fluid from the pool and form a lamella, or remain separate (separate ejecta sheet, SES); the transition between these regimes is associated with a fixed Capillary number, $\Ca=0.2$ \citep{Agbaglah2015}. Herein, we extend this knowledge to include pool movement and show that impact outcomes can differ upstream and downstream only when $\Ca<0.2$. The transition can be explained by considering the evolution of the ejecta sheet angle, giving a length-scale invariant parametrisation of $\Ca$ and the square root of the pool-droplet velocity ratio.

\section{Methods}

\subsection{Experiments}
\label{sec:exp}

\begin{table}
    \begin{center}
        \begin{tabular}{lccc}
            Fluid (\textbf{short name})         & Density, $\rho$ (\si{\kilogram\per\cubic\meter})  & Dyn. viscosity, $\mu$ (\si{\milli\pascal\second})  & Surface tension, $\sigma$ (\si{\milli\newton\per\meter})  \\[3pt]
            Distilled \textbf{water}                      & $997\pm1$                                         & $0.93\pm0.01$                                         & $72.4\pm0.2$      \\
            \textbf{21 vol\%} glycerol-water    & $1058\pm1$                                        & $1.9\pm0.0$                                           & $71.0\pm0.3$      \\
            \textbf{32 vol\%} glycerol-water    & $1089\pm1$                                        & $3.0\pm0.1$                                           & $70.3\pm0.2$      \\
            \textbf{43 vol\%} glycerol-water    & $1122\pm1$                                        & $5.4\pm0.2$                                           & $69.0\pm0.4$      \\
            \textbf{47 vol\%} glycerol-water    & $1134\pm2$                                        & $6.5\pm0.3$                                           & $68.4\pm0.5$      \\
            \textbf{51 vol\%} glycerol-water    & $1145\pm2$                                        & $8.0\pm0.3$                                           & $67.8\pm0.4$      \\
            \textbf{1 cSt} silicone oil         & $816$                                             & $0.8$                                                 & $17.4$            \\
            \textbf{2 cSt} silicone oil         & $870$                                             & $1.7$                                                 & $18.7$            \\
        \end{tabular}
        \captionsetup{width=1.0\linewidth}
        \caption{Fluid properties of the droplet and pool (same fluid). The density of each glycerol-water mixture (glycerol, Acros Organics 99\% pure) was measured using a \SI{25}{\milli\litre} density bottle and \SI{1}{\milli\gram} precision analytical balance, dynamic viscosity with a vibrational viscometer (Hydramotion Viscolite 700, \SI{0.1}{\milli\pascal\second} precision), and surface tension with a Sinterface BPA-2S tensiometer. Measured viscosities were confirmed to be consistent with known empirical correlations \citep{Cheng2008reading}. Silicone oils (Clearco Products) were used as received; the fluid properties reported are those from the product data sheet. All values are reported for $23\pm1\si{\degreeCelsius}$.\label{tab:fluids}}
    \end{center}
\end{table}

\begin{figure}
    \begin{center}
        \includegraphics{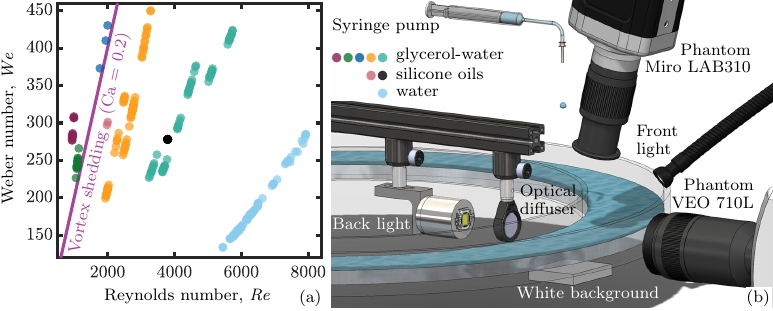}
    \end{center}
    \captionsetup{width=1.0\linewidth}
    \caption{(a)~$\We$ versus $\Rey$ for all experiments reported. The purple line delineates the known vortex shedding boundary \citep[$\Rey=5\We$, i.e. $\Ca=0.2$,][]{Agbaglah2015}. (b)~A rendering of the experimental setup.\label{fig:setup}}
\end{figure}

Millimetric droplets were dripped from a blunt end dispensing tip and impacted normally onto a deep pool, both consisting of the same fluid: distilled water, five glycerol-water mixtures, and two silicone oils -- see table~\ref{tab:fluids}. Six dispensing tips (15--30 gauge, outer diameters \SIrange{0.31}{1.83}{\milli\meter}) were used to modify the droplet diameter $D\in[2.20,3.75]~\si{\milli\meter}$, which is our characteristic length scale measured by the circle-fitted nominal radius, $\rn$ (see the supplementary material). The dispensing tip height relative to the pool free surface was varied (205--575~\si{\milli\meter}) to adjust the normal impact velocity, $\un\in[1.6,3.2]~\si{\meter\per\second}$. The variation in $\un$ was primarily responsible for the range of $\We=\rho \un^2D/\sigma\in[134,450]$, while the different fluids enabled considerable variation in $\Rey=\rho \un D/\mu\in[940,7930]$ (figure~\ref{fig:setup}a). As the Capillary number $\Ca = \We/\Rey = \mu\un/\sigma$ proves influential, we non-dimensionalise times with the visco-capillary time scale, $\mu D/\sigma$ (denoted $\tmustar$).

For all experiments, the pool depth $h$ was maintained such that $h/D>3$ (typically with $h\in[12,14]~\si{\milli\meter})$, which is sufficient that the pool can be considered deep for the early-time dynamics of interest here \citep{Slingshot,SykesJFM2023}. Pool movement was achieved using a belt-driven rotating table fitted with an optically clear annular tank (figure~\ref{fig:setup}b), whose dimensions are specified in the supplementary material. The inner walls were treated with a commercial hydrophobic coating (Rain-X Plastic Water Repellent) to curtail the formation of a meniscus that would interfere with side-view imaging. The experiments involving silicone oils used a similar setup described in \citet{DansPaper}.

Practical limitations restricted the maximum rotational frequency to around \SI{0.5}{\hertz}, with linear pool speeds at the impact point of $\ut\in[0,0.6]~\si{\meter\per\second}$ reported here. $\ut$ was calculated from the rotation rate of the table, which was determined by monitoring the time for each quarter rotation using an optical switch (Optek OPB900W55Z) connected to a data logger (Moku:Go). The implied $u_t$ therefore assumes that the fluid is in solid body rotation with the tank and that the effect of air drag on the free surface is negligible. Consequently, a spin-up time of approximately \SI{5}{\minute} was applied prior to all experiments, which is sufficient according to analysis of the typical timescales associated with the bottom-wall Ekman layer that is responsible for spin-up here \citep{Greenspan1963,HomiczGerber1987}. Particle tracking velocimetry experiments were also used to verify the actual free surface velocities -- see the supplementary material for both \citep{Crocker1996}.

Impacts were imaged from the side (through the tank side wall, figure~\ref{fig:setup}b) with a Phantom VEO 710L high-speed camera (\SIrange{7500}{14000}{fps}; \SIrange{3}{12}{\micro\second} exposure) in a shadowgraphy configuration, using a Laowa \SI{100}{\milli\meter} lens (\SIrange{64}{89}{\px\per\milli\meter}). A second front-lit high-speed camera (Phantom Miro LAB310; \SIrange{4800}{6300}{fps}; \SIrange{40}{70}{\micro\second} exposure), mounted \ang{15} from vertical to keep the path of the falling droplet free, simultaneously imaged from an oblique viewpoint using a Tamron SP AF \SI{90}{\milli\meter} lens (\SIrange{30}{40}{\px\per\milli\meter}).

\subsection{Direct numerical simulation}
    \label{sec:numerical}

\begin{figure}
    \begin{center}
        \includegraphics{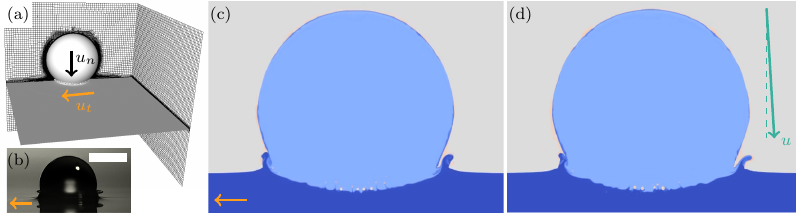}
    \end{center}
    \captionsetup{width=1.0\linewidth}
    \caption{(a) Computational box highlighting adaptive grid refinement. (b)~Experimental view of the case described by $\Ca = 0.105$ (32 vol\% fluid) and $\ut = \SI{0.15}{\meter\per\second}$ ($\un = \SI{2.45}{\meter\per\second}$, $\quantity = 0.25$), at $\tmustar = 2$. The upstream outcome is a lamella. The orange arrow indicates the direction of pool movement and the scale bar is \SI{2}{\milli\meter}. (c) The result of a simulation matching the conditions in panel~(b), with tracer fields used to visualise liquid originating from the droplet and the pool separately. As in the experiment, a lamella is seen upstream. (d) An equivalent oblique impact ($\uo = \sqrt{\ut^2 + \un^2} = \SI{2.455}{\meter\per\second}$; $\beta = \tan^{-1}(\ut/\un) = \ang{3.5}$) on a static pool ($\ut=\SI{0}{\meter\per\second}$). The droplet falls from left to right here, in the direction indicated by the green arrow (dashed is vertical). A lamella seen on the leading side, which corresponds to upstream on a moving pool.\label{fig:comp}}
\end{figure}

We constructed a high-fidelity computational counterpart of our system using the Basilisk open-source environment \citep{popinet2009accurate, popinet2015quadtree}, which has been successfully used in recent years for supporting both experimental and theoretical high-speed droplet impact research to gain additional physical insight into rapidly evolving ejecta sheet dynamics \citep{fudge2023drop, SykesJFM2023}. The computational box is three-dimensional, with a symmetry boundary condition used on the plane spanned by the pool and droplet velocity vectors (figure~\ref{fig:comp}(a)). The non-symmetry boundaries are $5\rn$ from the impact point, with either an imposed uniform unidirectional velocity field or outflow describing the remaining boundaries. Employing adaptive mesh refinement based on interfacial position location and changes in magnitude of velocity components and vorticity allows us to restrict the computational effort to $\mathcal{O}(10^7)$ computational grid cells while maintaining an $\mathcal{O}(1)~\si{\micro\meter}$ resolution level for the most resource-intensive flow regions, illustrated in figure~{\ref{fig:comp}(a)}. With these specifications, a typical run that represents approximately \SI{0.5}{\milli\second} in real time requires $\mathcal{O}(10^6)$ CPU hours to complete. A dedicated repository for the source code and typical parametric setup is provided on \href{https://github.com/OxfordFluidsLab/MovingPoolImpact/tree/main/DNS}{GitHub}. In section~\ref{sec:oblique}, we use these simulations to compare normal impact on a moving pool with oblique impact onto a static pool.

\section{Results and discussion}

\subsection{Impact outcomes on moving $\Ca<0.2$ pools}

\begin{figure}
    \includegraphics{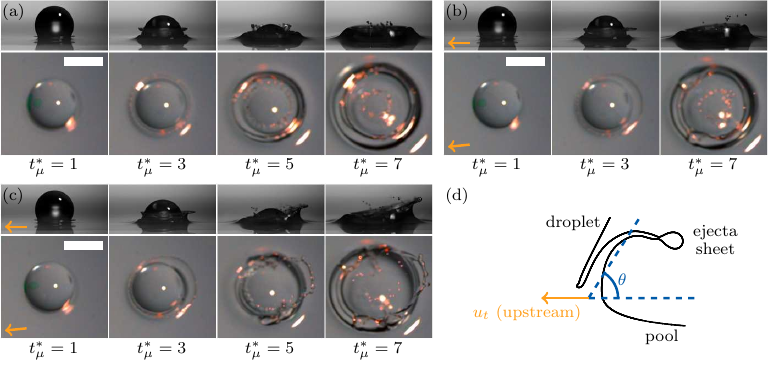}
    \centering
    \captionsetup{width=1.0\linewidth}
    \caption{(a)--(c) $\Ca=0.132$ ($\We=345$; $\un = \SI{3.10}{\milli\second}$) impact of a 32 vol\% droplet onto a 32 vol\% deep pool. (a) The pool is static: separate ejecta sheet, which is expected since $\Ca<0.2$ (figure~\ref{fig:setup}a). (b) The pool moves with $\ut = \SI{0.17}{\meter\per\second}$ ($\quantity=0.23$): separate ejecta sheet, but the ejecta sheet dynamics are not axisymmetric. (c) The pool moves with $\ut = \SI{0.26}{\meter\per\second}$ ($\quantity=0.29$): lamella upstream and a separate ejecta sheet downstream. (d) Sketch of an ejecta sheet, where $\theta$ is the ejecta sheet angle, as defined in \citet{Thoraval2012karman}. Orange arrows indicate the direction of pool movement and all scale bars are \SI{2}{\milli\meter}.\label{fig:mainmontage}}
\end{figure}

Initially, we consider $\Ca<0.2$ impacts, for which a static deep pool produces a SES outcome. Figure~\ref{fig:mainmontage} shows a representative case ($\Ca = 0.132$; 32~vol\%) where only the pool speed, $u_t \in \{0,0.17,0.26\}~\si{\meter\per\second}$ is varied. As expected, a generally axisymmetric SES is seen for a static pool in figure~\ref{fig:mainmontage}a. The ejecta sheet folding towards the axis of symmetry is clearly visible as an inner circle in the oblique view (bottom row), especially at $\tmustar=5$. For $\ut = \SI{0.17}{\meter\per\second}$ (figure~\ref{fig:mainmontage}b), the dynamics are not axisymmetric, but a SES outcome is maintained in all directions. However, when the pool speed is increased further to $\ut = \SI{0.26}{\meter\per\second}$ (figure~\ref{fig:mainmontage}c), the ejecta sheet develops into a lamella upstream. This outcome is typically seen on viscous ($\Ca>0.2$) or shallow \citep{SykesJFM2023} static pools. A SES is maintained downstream and in most directions (see the oblique view) except for $\sim\ang{70}$ around the upstream side. That is, above a critical pool speed, the upstream impact outcome transitions from a SES to a lamella.

\subsection{Delineating the upstream transition boundary}
\label{sec:delineate}

The Capillary number is the appropriate quantity to represent droplet impact on deep pools \citep{Agbaglah2015}. To delineate the upstream transition, we also need to determine a suitable dimensionless quantity to represent pool movement. \citet{Slingshot} developed a simple geometric model to describe ejecta sheet dynamics, in which the ejection velocity is directed tangentially to a sphere approximating the droplet and is proportional in magnitude to $u_n\cos\theta$. Here, $\theta$ is the angle between the horizontal and the normal to the ejecta sheet base \citep[carefully defined in][]{Thoraval2012karman}, as sketched in figure~\ref{fig:mainmontage}d. Geometrically, $\theta$ is also the angle subtended at the sphere/droplet centre by the droplet impact velocity vector (vertical in the case of normal impact) and radial line to the ejecta sheet base \citep[see figure~8a inset in][]{Slingshot}. In our work, the pre-impact droplet and pool velocity vectors are orthogonal; the resultant velocity vector is directed $\beta$ from the vertical, where $\tan\beta = \ut/\un \approx \beta$ since $\ut\ll\un$. For a geometrically-equivalent oblique impact at an angle $\beta$ from the vertical on a static pool (discussed further in section~\ref{sec:oblique}), the ``leading side'' \cite[adopting the nomenclature of][]{Reijers2019} is equivalent to upstream, which can be understood by considering that a normally-impacting droplet would appear to be falling backwards by an observer travelling on a moving pool. Therefore, for a $\beta$ oblique impact, $\theta$ is effectively reduced, as the droplet impact velocity vector is no longer vertical -- it is pointed towards the leading side. \citet{Thoraval2012karman} found that $\theta$ increases as $\theta \sim \sqrt{\Rey} \propto \un^{1/2}$ for a static deep pool. Here $\ut/\un \approx \beta$, where $\beta$ is a constant for a given impact, and therefore suggests that $\ut$ may act inversely proportional to $\un$ in relation to the evolution of $\theta$ upstream on a moving pool, i.e. $\theta^{-1} \sim \ut^{1/2} \propto \sqrt{\Reyt} = \sqrt{\rho\ut D/\mu}$. That is, pool movement constrains the growth of $\theta$ upstream, although this is practically unmeasurable from our experiments. However, the same conclusion can be reached mechanistically by considering the ejecta sheet base moves with the pool whilst the ejecta sheet evolves, which for an otherwise static ejecta sheet would reduce $\theta$ as sketched in figure~\ref{fig:mainmontage}d. This analysis hints at the importance of the non-dimensional quantity $\sqrt{\Reyt/\Rey}=\quantity$.

\begin{figure}
    \includegraphics{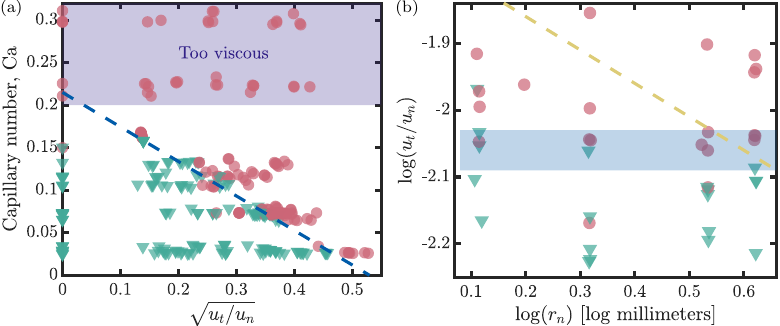}
    \captionsetup{width=1.0\linewidth}
    \caption{Upstream impact outcomes for normal droplet impact on a moving deep pool. Red circular markers indicate a lamella, while green triangular markers indicate a separate ejecta sheet. (a)~This regime map includes all experimental conditions described in section~\ref{sec:exp}: $\We\in[134,450]$ and $\Rey\in[940,7930]$. The same regime map with markers coloured by the fluid involved is provided as supplementary material. The blue dashed line delineates a linear least squares fit to the transition. (b)~$\Ca = 0.072\pm0.002$ (21 vol\% and 1~cSt fluids) with $r_n\in[1.11,1.87]~\si{\milli\meter}$ to assess the influence of length scale. The blue patch indicates the approximate upstream transition identified. For comparison, the yellow dashed line indicates a $\sfrac{1}{4}$ exponent (an arbitrary example of a weak dependence for demonstrative purposes) dependence on $r_n$ , i.e. $r_n^{1/4}\quantity$ for fixed $\Ca$.\label{fig:rm}}
\end{figure}

Figure~\ref{fig:rm}a shows a regime map of $\Ca$ against $\quantity$, with each data point colour and shape indicating the observed upstream impact outcome (green triangle for SES; red circle for lamella). This regime map contains \emph{all} of the experimental conditions described in section~\ref{sec:exp}, including a wide range of fluid properties (see figure~S2 in the supplementary material), droplet diameters, and pool \& droplet velocities. Together, $\Ca$ and $\quantity$ near perfectly separate all upstream impact outcome types, producing a sharp boundary for $\Ca<0.2$. Note that there is no upstream transition (for all studied pool speeds) when $\Ca>0.2$, which is expected as the impact outcome is already a lamella on a static pool.

Visually, the $\Ca<0.2$ transition boundary appears linear, so it is tempting to attempt a least-squares linear fit. Practically, we do this for all the data (except silicone oils) for which $\Ca<0.15$, in $\Ca$ bins of 0.01, fitting to the midpoint of the first lamella and last SES outcome (when increasing pool speed), the extent of which is commensurate to the error in determining $\quantity$. The result is the blue dashed line in figure~\ref{fig:rm}a, which is consistent with the $\Ca\approx0.16\pm0.01$ silicone oil data. Most notably, the fit approximately recovers the known $\Ca=0.2$ static deep pool threshold \citep{Agbaglah2015}, so a single constant (representing the gradient, $\approx-0.4$) delineates the upstream transition alongside $\Ca$ and $\quantity$. Intriguingly, neither of these two quantities contain a length scale, despite the experiments in figure~\ref{fig:rm}a having $D\in[2.20,3.75]~\si{\milli\meter}$. This observation suggests that the upstream transition is length-scale invariant, which we now investigate.

\subsection{Length-scale invariance of the upstream transition}

To assess the involvement of the length scale in the upstream transition, we consider experiments with a fixed $\Ca$ across the full range of $D=2r_n$ studied. We suppose that $r_n^\alpha\sqrt{u_t/u_n} = c$, where $c$ is a constant, and $\alpha$ is an unknown exponent to be determined. Hence, $\log(\ut/\un) = -2\alpha\log(r_n) + 2\log(c)$, so $-2\alpha$ would be the upstream transition gradient on a graph of $\log(\ut/\un)$ against $\log(r_n)$ for fixed \Ca. Such a graph (for fixed $\Ca = 0.072\pm0.002$; 21 vol\% and 1~cSt fluids) is shown in figure~\ref{fig:rm}b. Within the expected experimental error, the upstream transition appears flat (indicated within the blue patch), suggesting that $\alpha=0$. By way of comparison, the orange dashed line delineates the expected transition for a relatively weak length-scale dependence of $\alpha=0.25$ (an arbitrary choice), which is fixed at the bottom of the blue patch on the right-hand side. This analysis strongly suggests that the upstream transition is length-scale invariant.

\subsection{Ejecta sheet dynamics}

\begin{figure}
    \includegraphics{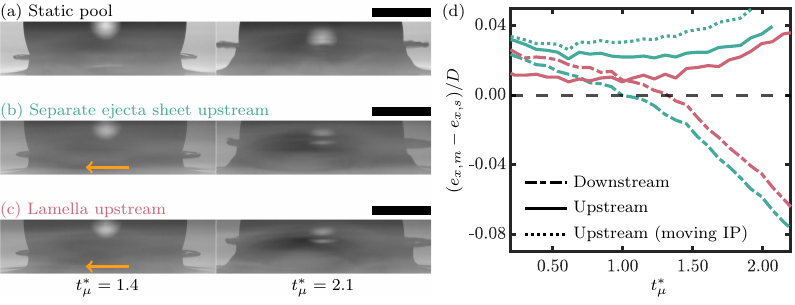}
    \captionsetup{width=1.0\linewidth}
    \caption{(a)--(c) High resolution ($\SI{327}{\px\per\milli\meter}$) images of the early-time ejecta sheet dynamics of $\Ca~=~0.115\pm0.002$ impact (32 vol\% fluid). (a) Static pool. (b)--(c) $\ut=0.20\pm0.01~\si{\meter\per\second}$ ($\quantity=0.27$): (b) separate ejecta sheet upstream; (c) lamella upstream. Orange arrows indicate the direction of pool movement and all scale bars are \SI{1}{\milli\meter}. (d) Data indicating the difference between the horizontal extent of the ejecta sheet on moving and static pools, from the experiments in panels (a)--(c). $e_{x,p}$ is the horizontal position of the ejecta sheet tip from the original impact point (IP); $p=m$ for a moving pool (either (b) green or (c) red) and $p=s$ for a static pool. In one case (`moving IP', dashed line), the IP is translated with the moving pool speed over time.\label{fig:stretch}}
\end{figure}

To attain a detailed view of the early-time ejecta sheet dynamics, the side-view camera was replaced by a Phantom TMX~5010 (\SI{110000}{fps}; \SI{1.0}{\micro\second} exposure) equipped with a Navitar 12X Zoom lens (for \SI{327}{\px\per\milli\meter}) for the experiments in figure~\ref{fig:stretch}. Here, $\Ca = 0.115 \pm 0.002$ (32 vol\% fluid), with a SES outcome on a static pool (figure~\ref{fig:stretch}a). For both figures~\ref{fig:stretch}b and \ref{fig:stretch}c, $\ut$ was set such that $\quantity = 0.27$, which is approximately the upstream transition threshold according to figure~\ref{fig:rm}a. Figure \ref{fig:stretch}b shows a SES upstream (i.e. before the transition); figure \ref{fig:stretch}c a lamella upstream (i.e. after the transition).

Qualitatively, the ejecta sheets appear similar at $\tmustar=1.4$, whether the pool is static or moving. However, the vortex separation process that produces a SES outcome is known to start at $\tmustar=1.05$ \citep{Agbaglah2015}, which must be inhibited upstream in figure~\ref{fig:stretch}c to generate the lamella outcome. It appears that the ejecta sheet is slightly thicker on the moving pools in figure~\ref{fig:stretch}, although this cannot be systematically confirmed within reasonable error tolerances with our current experimental and numerical setups. By $\tmustar=2.1$, the impact outcomes are visually evident and the ejecta sheet has notably different horizontal extents from the impact point, $e_{x,p}$ ($p=m$ moving; $p=s$ static) upstream and downstream. These observations are confirmed quantitatively in figure~\ref{fig:stretch}d. Around the aforementioned $\tmustar=1.05$, the ejecta sheet considerably shortens downstream, relative to a static pool (so $e_{x,m}<e_{x,s}$), when the pool is moving. Upstream, the ejecta sheet is always stretched (so $e_{x,m}>e_{x,s}$), even when measuring $e_{x,m}$ from the fixed impact point (dash-dot and solid lines). However, pool movement translates the impact point downstream, which conceivably augments ejecta sheet stretching upstream. To account for this effect, $e_{x,m}$ can be measured from the moving impact point (practically, adding \SI{24}{\micro\meter} per $\tmustar$ time unit to $e_{x,m} - e_{x,s}$), which is shown in figure~\ref{fig:stretch}d as a dotted line for the upstream side of the figure~\ref{fig:stretch}b data. These analyses indicate that the ejecta sheet is considerably stretched upstream on a moving pool, which raises the prospect that pool movement thins the ejecta sheet at its base and restricts flow into it from the pool. It is this mechanism that we found to yield a lamella on sufficiently shallow pools in our previous work \citep{SykesJFM2023}, where ejecta sheet stretching was found to be caused by a pressure confinement effect of the pool base. Hence, similar physical mechanisms may be at play in both configurations.

\subsection{Downstream impact outcome}

\begin{figure}
    \includegraphics{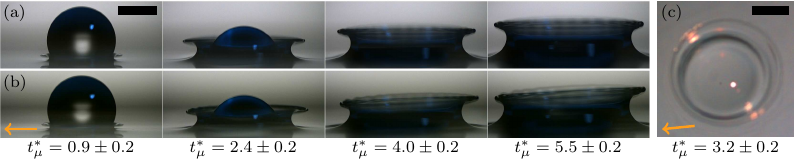}
    \captionsetup{width=1.0\linewidth}
    \caption{$\Ca = 0.213\pm0.002$ with the 43 vol\% fluid. (a) Static pool. (b)--(c) Moving pool with $\quantity = 0.33$. The blue tinge in (a) and (b) is an artifact of repeated inner wall surface treatments and the light source. Orange arrows indicate the direction of pool movement and all scale bars are \SI{1}{\milli\meter}.\label{fig:viscous}}
\end{figure}

In so far as pool movement effectively constraints the evolution of $\theta$, the opposite is true downstream. Therefore, were the root cause of the upstream transition purely geometric according to section~\ref{sec:delineate}, we might expect pool movement to recover a SES \emph{downstream} when the static pool impact outcome is a lamella ($\Ca > 0.2$). Throughout our experimental campaign, no such transition was seen. Figure~\ref{fig:viscous} exemplifies typical dynamics for $Ca>0.2$ on static (figure~\ref{fig:viscous}a) and relatively fast-moving (figures~\ref{fig:viscous}b and \ref{fig:viscous}c, $\ut = \SI{0.29}{\meter\per\second}$) pools. It is notable that little asymmetry can be seen in the side views (e.g. figure~\ref{fig:viscous}b), especially at early times. Moreover, the oblique view (figure~\ref{fig:viscous}c) shows remarkably smooth dynamics and offers no hint of an instability that is typically apparent with the SES outcome.

Within the limits of the experimental design (namely the limitation on pool speed), our results indicate that there is no lamella to SES transition downstream for $\Ca>0.2$. That is, when the pool is ``too viscous'' (figure~\ref{fig:rm}a), there is no other outcome than a lamella when the pool boundary layer is negligible. The lack of ``reversibility'' in the SES-lamella transition hints at a fundamental idea relevant to both moving and static pools: the cause of the transition is not purely geometric. Rather, the association of the transition to $\tmustar=1.05$ \citep[first made by][]{Agbaglah2015} appears causational, hinting that the root cause of a SES is an instability at the base of the ejecta sheet that enables vortex shedding. For $\Ca>0.2$, viscosity suppresses the instability and the ability to attain the SES outcome.

\subsection{Oblique impact}
\label{sec:oblique}

Except the presence of a boundary layer on the moving pool, prior to coalescence normal droplet impact onto a moving pool is geometrically equivalent to oblique impact on a static pool at an angle $\beta$ from vertical and a velocity $\uo$, where $\un = \uo\cos\beta$ and $\ut = \uo\sin\beta$. On solid surfaces, the two types of impact are equivalent regarding the resulting spreading and splashing behaviour \citep{Buksh2020}. The boundary layer on moving pools is known to have qualitative effects on impact outcomes when $\ut\gtrsim\un$, and can even support the weight of an impacting droplet so that it `surfs' without coalescing \citep{Castrejon2016}. Of course, the persistent effect of pool movement does affect longer timescale dynamics, such as bouncing, compared to an oblique impact on a static pool \citep{DansPaper}. However, when $\un\ll\ut$ we hypothesise that the boundary layer does not play a decisive role on the ejecta sheet dynamics, so we expect a transition on the leading side (equivalent to upstream on a moving pool, see section~\ref{sec:delineate}) of a sufficiently oblique impact.

Previously, oblique impacts of small droplets ($\si{115}\pm\SI{15}{\micro\meter}$) have been achieved via deflection with an electric field, which is a process used in some commercial continuous inkjet printers. The resulting long timescale dynamics (e.g. crown shape and Worthington jet inclination) appear qualitatively similar to our asymmetric moving pool experiments \citep{GielenPRF2017}. However, the greater inertia of larger droplets in our study makes highly-controlled oblique impacts challenging to engender experimentally, so we turn to simulations that have been used to successfully study oblique impacts previously \citep{cimpeanu2018three}. We reproduced the experiment in figure~\ref{fig:comp}b ($\Ca = 0.105$, 32 vol\%, $\un=\SI{2.45}{\meter\per\second}$, $\ut=\SI{0.15}{\meter\per\second}$) computationally in figure~\ref{fig:comp}c, with good qualitative agreement. The equivalent oblique impact has $\uo=\SI{2.455}{\meter\per\second}$ (to 3 decimal places) and $\beta=\ang{3.5}$. This very small angle is shown diagramatically (as green arrows) inset in figure~\ref{fig:comp}d, alongside the results of the equivalent oblique impact simulation. As predicted, a lamella is formed on the leading side rather than the SES outcome, supporting our rationale that the moving pool boundary layer does not play a decisive role. This result indicates that our findings for moving pools are generalisable to equivalent oblique impacts.

\section{Conclusions}

For $\Ca<0.2$ normal impacts on deep pools, the typical separate ejecta sheet dynamics transition to a lamella only upstream, above a critical pool speed. Such asymmetric outcomes arise with little pool movement (e.g. above a pool-droplet velocity ratio of $\ut/\un = 0.08$ when $\Ca = 0.10$), raising the prospect that the motion of rivers and oceans may result in asymmetric raindrop impact. The transition boundary is well-parametrised by $\Ca$ and $\quantity$, which are both independent of length scale. By considering experiments with a fixed $\Ca$, we concluded that the upstream transition is length-scale invariant. We explained the importance of the $\quantity$ physically as a constraining effect of pool movement on the ejecta sheet angle evolution upstream. A linear fit to the transition boundary recovers the known $\Ca=0.2$ threshold on static deep pools \citep{Agbaglah2015}. No reverse transition (lamella to separate ejecta sheet) is seen downstream for $\Ca>0.2$, which suggests that the underlying mechanism causing the separate ejecta sheet outcome is not purely geometric. Instead, our results suggest an instability (characterised by the visco-capillary time scale) that is suppressed by high viscosity; this conclusion applies to any like-fluid pool impact. Our 3D DNS framework enabled us to confirm similar dynamics for an equivalent oblique impact on a static pool, suggesting that the moving pool boundary layer does not have a decisive role on impact outcomes when $\ut\ll\un$. Our findings offer insight into common natural and industrial scenarios (e.g. ocean air-sea exchange, inkjet printing) involving moving pools where satellite droplet production is often an important consideration, as well as contributing to the fundamental understanding of the physical mechanisms that determine impact outcomes on pools more generally.\\

\begin{appen}

    \noindent {\bf Supplementary Material.} Supplementary material are available at [URL to be inserted by publisher].

    \noindent {\bf Acknowledgments.} We thank Prof. Peter Read (Oxford Physics) for lending the rotating table base, Duncan Constable (Oxford Engineering) for helping to manufacture the annular tank, and Oliver Sand for setting up the rotating table at Brown.

    \noindent {\bf Funding.} This work was funded by the US National Science Foundation (NSF CBET-2123371), UK Engineering and Physical Sciences Research Council (EP/W016036/1), and the John Fell Fund (0014320).

    \noindent {\bf Data availability.} Data are available at \url{https://github.com/OxfordFluidsLab/MovingPoolImpact}.

    \noindent {\bf Declaration of interests.} The authors report no conflict of interest.

\end{appen}

\bibliographystyle{jfm}
\bibliography{main}

\end{document}